\documentclass[cup6b]{cupbook}
\usepackage{graphicx}
\usepackage{natbib}
\title[Rome, Italy, 27--30 April 2009]
      {The coming of age of X-ray polarimetry}
\author{}
\date{}
\begin{document}
\pagenumbering{arabic}
%%%%%%%%%%%%%%%%%%%%%%%%%%%%%%%%%%%%%%%%%%%%%%%%%%%%

% YOUR CONTRIBUTION GOES HERE

% Insert here the author list and the affiliation
\author[McNamara et al.]{Aimee McNamara (University of Sydney) \and Zdenka Kuncic (University of Sydney) 
  \and Kinwah Wu (University College London)}
% Title of the contribution
\chapter{Polarization of Compton X-rays from Jets in AGN}

% Abstract
\abstract{ We investigate the polarization of Compton scattered X-rays from relativistic jets 
   in active galactic nuclei (AGN) 
      using Monte Carlo simulations. 
   We consider three scenarios: 
      scattering of photons from an accretion disk, scattering of cosmic microwave background (CMB) photons, 
      and synchrotron self-Comptonization (SSC) within the jet.   
   For Comptonization of thermal disk photons or CMB photons     
      the maximum linear polarization attained is slightly over 20\% at viewing angles close to 90$^\circ$ 
   The value decreases with the viewing inclination. 
   For SSC, the maximum value may exceed 80\%. 
   The angle dependence is complicated, and it varies with the photon injection sites.  
    Our study demonstrates that X-ray polarization, in addition to multi-wavelength spectra,  
       can distinguish certain models for emission and particle acceleration in relativistic jets.         }

\section{Introduction}

Observations of extended jets in AGN by {\it Chandra} have revealed 
  that the origins of their X-ray emission is less trivial than previous thought 
  \citep[see][for X-ray jet surveys]{Marshall05,Sambruna04}. 
The X-rays may arise from various processes. 
The polarization in the radio and optical bands 
  suggest that the emission is generated by synchrotron process \citep{Jorstad07}. 
Thus, synchrotron and synchrotron self-Compton (SSC) emission 
  are candidates for the X-ray continuum emission \citep{Maraschi92}. 
However, the X-rays can also be generated from external Comptonization (EC) 
   of disk blackbody radiation \citep{Wagner95} or of the CMB \citep{Celotti01}. 
It has been suggested that X-ray polarization measurements 
  are able to discriminate these competing emission mechanisms. 
To date only a few approximate analytical predictions 
  have been made for SSC \citep{Bjornsson82b, Begelman87, Celotti94} 
  or EC polarizations  \citep{Poutanen94}. 
Here we show results of our X-ray polarization calculations  
  for photons scattered by energetic electrons in jets at relativistic bulk speeds. 
We consider Compton scattering of thermal photons emitted from an underlying accretion disk 
  and the scattering of polarized synchrotron photons emitted within the jet. 
We also consider Compton scattering of CMB photons by the bulk flow of jet electrons. 

%%%%%%%%%%%%%%%%%%%%% 
\begin{figure}
\centering
\vspace*{0.4cm}
\includegraphics[scale=0.24]{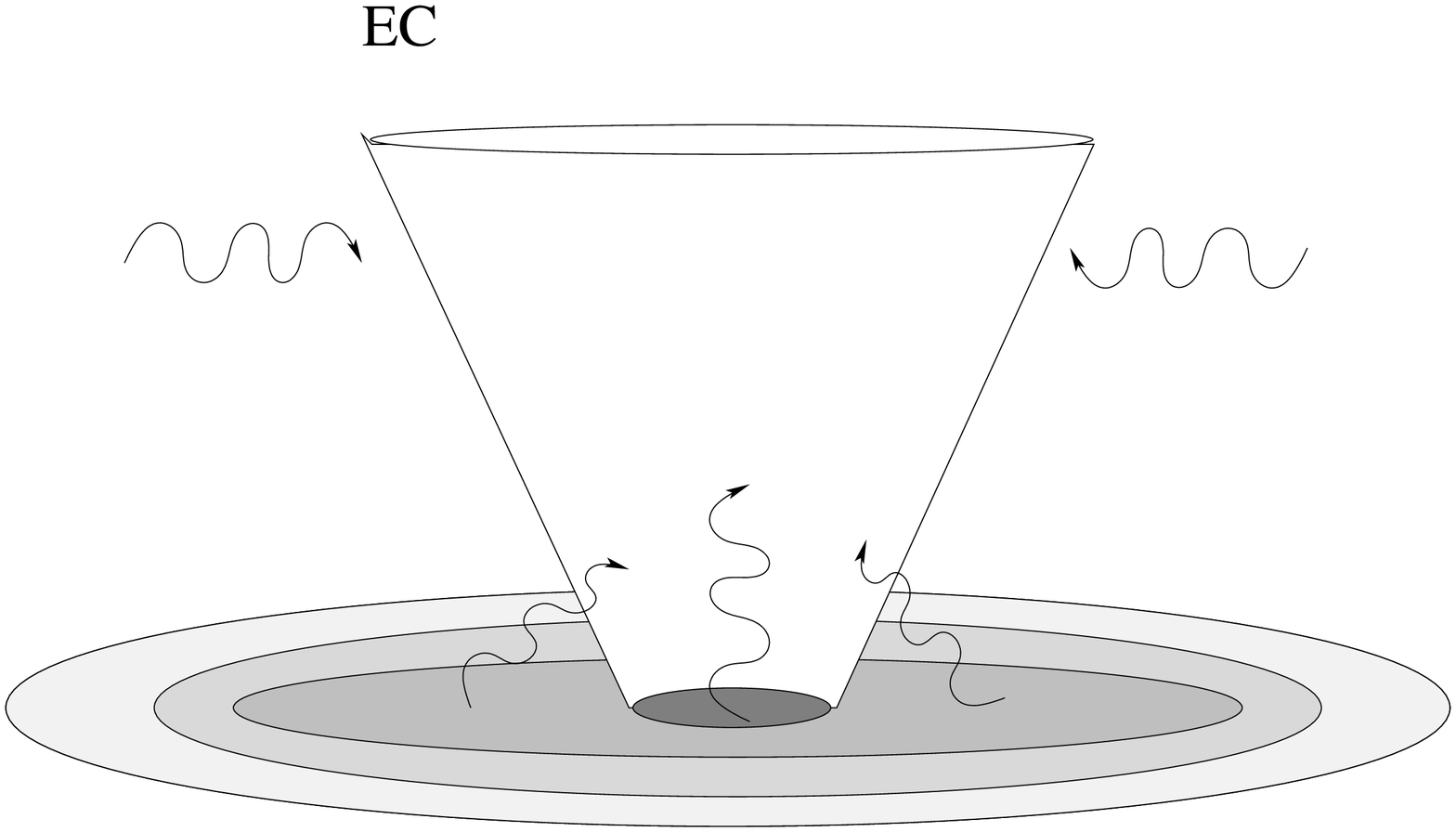} 
 \hspace*{0.3cm}
\includegraphics[scale=0.24]{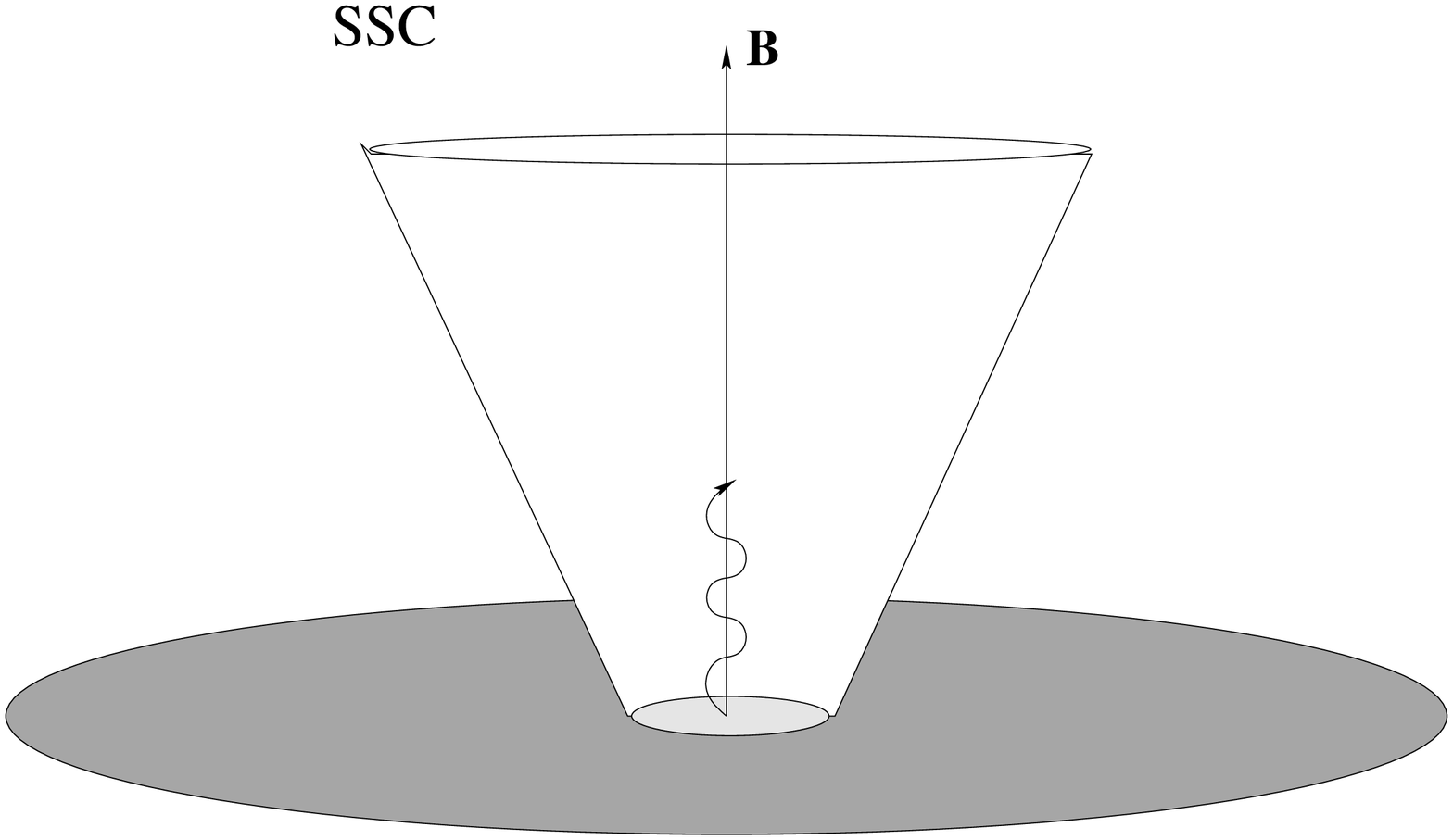}
\caption{Schematic illustrations of the geometrical set up of the calculations 
   for the cases of external Comptonization (EC) (left) and SSC (right). 
  For the EC case thermal photons are either injected from an accretion disk 
     or injected isotropically around the conical jet. 
  For the SSC case, linearly polarized synchrotron photons are injected at specific locations 
     inside the jet. 
  A magnetic field \textbf{B}  is permeated in the jet along the bulk flow direction.  }
\label{jet_geo}
\end{figure}

\section{Models} 

We consider a conical jet with a bulk Lorentz factor $1 < \Gamma_{\rm j} < 10$, 
   launched at a height $z_{\rm 0}$ above the disk midplane (Figure~\ref{jet_geo}).   
The jet base has a radius  $r_{\rm b}$, and the jet power is  $P_{\rm j}$.    
The electrons have a number density distribution, $N_{\rm e} (\gamma) = K\gamma^{-p}$, 
   where $\gamma$ is the Lorentz factor,  $p$ is the eneregy spectral index,  
   and $K$ is obtained from  $N_{\rm e} = \int_{\gamma_{\rm min}}^{\gamma_{\rm max}} N_{\rm e} (\gamma) \, d\gamma$. 
Along the jet $N_{\rm e}$ falls off according to  
  $N_{\rm e}(z) = N_{\rm 0} ({z}/{z_{\rm 0}})^{-2}$, where 
\begin{equation}
	N_{\rm 0} = N_{\rm e}(z_{\rm 0}) 
	 \approx \frac{P_{\rm j}}{ \beta_{\rm j}   \Gamma_{\rm j} (\Gamma_{\rm j} -1)m_{\rm p}c^3\pi r_{\rm b}^2} \  . 
\end{equation}

Polarized synchrotron photons are emitted isotropically 
  by the electrons with Lorentz factors $\gamma_{min} \leq \gamma \leq \gamma_{\rm max}$ 
  in the plasma rest frame. 
The emission is Lorentz boosted by the relativistic bulk flow of the jet. 
The accretion disk emission is thermal, with the photons generated according to the flux  
\begin{equation}
	F(R) = \frac{3GM\dot M_{\rm a}}{8\pi R^3} \left[1 - \left(\frac{R}{R_{\rm i}}\right)^{-1/2}\right]  \  , 
\label{Flux}	
\end{equation}	  
   where $R_{\rm i}$ is the inner disk radius and ${\dot M}_{\rm a}$ is mass accretion rate of the disk. 
The CMB photons are generated from a single-temperature blackbody 
  with $T_{\rm bb} = 2.8 \, (1+z)$~K. 
They are injected isotropically and uniformly around the jet. 
In this work, we set the redshift $z=2$. 
The numerical values of the model parameters are the same as those in \citep{McNamara09}. 
 
The non-linear Monte-Carlo algorithm given in \citep{Cullen01a, Cullen01b} 
  is used to determine the scattering event and the photon transport in the jet.  
The polarization is calculated in each scattering event, 
  and the emergent photons are summed,  
  following the procedures as described in  \citep{McNamara08a, McNamara08b}. 
In the simulations, three different co-ordinate frames  
  -- the electron rest frame,  the plasma rest frame and the observer's frame -- are used. 
The transformation of the polarization of the photons between the frames is given in  \citep{McNamara09}.   

%%%%%%%%%%%%%%%%%%%%%% 
\begin{figure}
\centering
\includegraphics[scale=0.4]{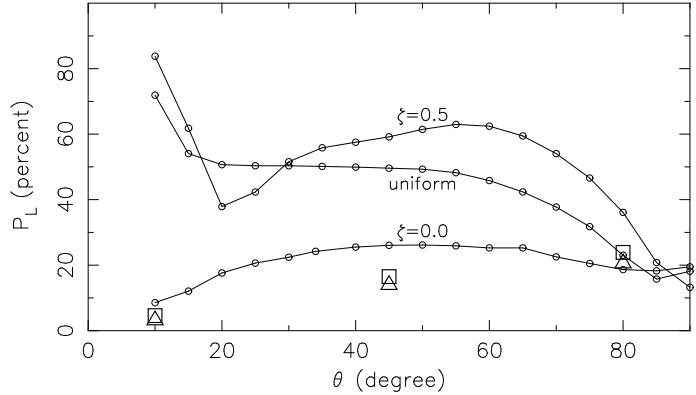}
\caption{
  Linear polarization $P_{\rm L}$ of Compton scattered photons 
     with energies between $1-10$~keV 
    as a function of the viewing inclination angle $\theta$. 
  The solid lines corresponding to the SSC cases 
     with seed photons injected  at the jet base ($\zeta =0$), in the middle of the jet ($\zeta = 0.5$) 
     and uniformly along the jet. 
  The triangles correspond to the EC case with seed photons generated from an accretion disk. 
  The squares correspond to the EC case with CMB photon injection.  
  }
\label{am_jet}
\end{figure}

\section{Results and discussions} 

The results of our simulations are shown in Figure~\ref{am_jet}. 
For the EC case with photons injected from an accretion disk, 
  the degree of linear polarization $P_{\rm L}$ increases 
  with the viewing inclination angle $\theta$.   
At $\theta = 10^\circ$, $P_{\rm L} \approx 3\%$, 
  and at  $\theta = 80^\circ$, $P_{\rm L} \approx 21\%$. 
The angle dependence of the polarization for the case with CMB photon injection   
  is very similar,   
  with $P_{\rm L} \approx 3\%$ at $\theta = 10^\circ$,  
  increasing to 24\% at at $\theta = 80^\circ$.  
In these two cases, the average polarization of the seed photons is zero.   
Large linear polarization results when large-angle scattering events dominate. 
As the jet plasma has bulk relativistic motion,  
  the incoming photons are essentially seen as headon in the electron rest frame. 
Regardless of the initial angle distribution of the seed photons, 
  most photons emerging at large $\theta$ 
  have undergone a large scattering event immediately before escaping from the jet. 
Thus, $P_{\rm L}$ increases with $\theta$, 
  and the degrees of polarization are similar for the two EC cases 
  with thermal photon injection.    

The polarization is substantially higher in the SSC cases than in the EC cases 
  -- above the 20\% level, except 
  when the photons are injected at the jet base and when the viewing inclination is low (small $\theta$).  
It reaches 70\% or higher if photon injection occurs at substantial heights or uniformly along the jet.  
The high polarization is partially due to the fact that the seed photons are polarized. 
Scattering randomizes the photon polarization angles.   
Multiple scattering does not always increase the net average polarization, 
  thus the angle dependence is more complicated in the SSC cases than the in the EC cases, 
  where net polarization is generated by multiple scatterings of unpolarized thermal photons.   
The polarization values are similar for all cases when $\theta$ is close to 90$^\circ$.  
This is due to the fact that  
  the corresponding incident photons have similar properties, in the rest frame of the electrons,  
  at the last scattering before they escape from the jet. 

Our study has shown that X-ray polarization can distinguish 
   whether the X-rays are generated from EC or from SSC. 
It can also discriminate 
  between the particle acceleration sites in the case of SSC emission.

%\subsection{A subsection}
%\subsubsection{A subsubsection}
%\theendnotes

%%%%%%%%%%%%%%%%%%%%%%%%%%%%%%%%%%%%%%%%%%%%%%%%%%%%
\end{document}